\documentclass[twocolumn,letterpaper,aps,prd,superscriptaddress,showpacs,nofootinbib,floatfix]{revtex4}
\usepackage{graphicx}
\usepackage{subfigure}
\usepackage{slashbox}
\usepackage{amsmath}
\usepackage{brackets}
\usepackage{psfrag}
\begin{document}

\ProvideTextCommandDefault{\textonehalf}{${}^1\!/\!{}_2\ $}

\title{Laboratory Search for Spin-dependent Short-range Force from Axion-Like-Particles using Optically Polarized $^{3}$He gas}

\affiliation{Triangle Universities Nuclear Laboratory and Department of Physics, Duke University, Durham, North Carolina 27708, USA}
\affiliation{Department of Physics, Indiana University, Bloomington, Indiana 47408, USA}
\affiliation{Department of Physics, Shanghai Jiaotong University, Shanghai, 200240, China}

\author{P.-H.~Chu}
\affiliation{Triangle Universities Nuclear Laboratory and Department of Physics, Duke University, Durham, North Carolina 27708, USA}
\author{A.~Dennis}
\affiliation{Department of Physics, Indiana University, Bloomington, Indiana 47408, USA}
\author{C.~B.~Fu}
\affiliation{Department of Physics, Shanghai Jiaotong University, Shanghai, 200240, China}
\author{H.~Gao}
\affiliation{Triangle Universities Nuclear Laboratory and Department of Physics, Duke University, Durham, North Carolina 27708, USA}
\author{R.~Khatiwada}
\affiliation{Department of Physics, Indiana University, Bloomington, Indiana 47408, USA}
\author{G.~Laskaris}
\affiliation{Triangle Universities Nuclear Laboratory and Department of Physics, Duke University, Durham, North Carolina 27708, USA}
\author{K.~Li}
\affiliation{Department of Physics, Indiana University, Bloomington, Indiana 47408, USA}
\author{E.~Smith}
\affiliation{Department of Physics, Indiana University, Bloomington, Indiana 47408, USA}
\author{W.~M.~Snow}
\affiliation{Department of Physics, Indiana University, Bloomington, Indiana 47408, USA}
\author{H.~Yan}
\affiliation{Department of Physics, Indiana University, Bloomington, Indiana 47408, USA}
\author{W.~Zheng}
\affiliation{Triangle Universities Nuclear Laboratory and Department of Physics, Duke University, Durham, North Carolina 27708, USA}
\date{\today}

\begin{abstract}
The possible existence of short-range forces between unpolarized and polarized spin-$\frac{1}{2}$ particles has attracted the attention of physicists for decades. These forces are predicted in various theories and provide a possible new source for parity (P) and time reversal (T) symmetry violation.  We use an ensemble of polarized $^3$He gas in a cell with a 250 $\mu$m thickness glass window to search for a force from pseudoscalar boson exchange over a sub-millimeter ranges. This interaction would produce a NMR frequency shift as an unpolarized mass is moved near and far from the polarized ensemble. We report a new upper bound with a factor of 10-30 improvement on the product $g_{s}g_{p}^{n}$ of the scalar couplings to the fermions in the unpolarized mass, and the pseudoscalar coupling of the polarized neutron in the $^{3}$He nucleus for force ranges from $10^{-4}$ to $10^{-2}$ m, which corresponds to a mass range of $2\times10^{-3}$ to $2\times10^{-5}$ eV for the pseudoscalar boson. This represents the most sensitive search that sets a direct limit in the important ``axion window". 
\end{abstract}

\maketitle
Possible short-range forces between unpolarized and polarized spin-$\frac{1}{2}$ particles can provide a new source for parity ($P$) and time reversal ($T$) symmetry violation~\cite{Leitner1964}. Moody and Wilczek~\cite{Moody1984} proposed a force from the exchange of spin-0 bosons which can couple to fermions through scalar and pseudoscalar vertices. The scalar coupling is spin-independent and depends only on the fermion density. The pseudoscalar coupling is entirely spin-dependent. The resulting spin-dependent short-range force (SDSRF) has a Yukawa-type interaction potential from one boson exchange of the form
\begin{align}
V(r) = \frac{g_s g_p \hbar^2 }{8\pi m_p} (\hat{\sigma}\cdot\hat{r})(\frac{1}{r\lambda}+\frac{1}{r^2})\exp(-r/\lambda)
\label{eq:potential}
\end{align}
where  $\hat{r}$ is the unit vector from the unpolarized particle to the polarized particle, $\hat{\sigma}$ is the spin of the polarized particle, $m_p$ is the polarized particle mass, $g_s g_p$ is the product of couplings of the scalar vertex in the unpolarized matter and the pseudoscalar vertex of the polarized particle, and $\lambda$ is the force range. Such forces may be induced by pseudoscalar bosons like the axion~\cite{Peccei1977}, axion-like-particles (ALPs)~\cite{Jaeckel2010} or a very light spin-1 boson~\cite{Fayet1996} which are candidates for cold dark matter~\cite{Kolb1990}. Current experimental and astrophysical observation restricts the axion mass between 1 $\mu$eV to 1 meV, corresponding to a force range between 2 cm to 20 $\mu$m, the so-called ``axion window"~\cite{Antoniadis2011}. Many ALPs, predicted by string theory~\cite{Svrcek2006} and many extensions to the standard model~\cite{Dobrescu2006,Adelberger2009}, also predict weak forces in this range. 
\vspace{-0.02in}

Several experiments have been performed to search for SDSRF using different techniques. Some examples include the torsion pendulum~\cite{Ritter1993,Hammond2007,Hoedl2011}, neutron bound states on a mirror in the Earth's gravitational field~\cite{Baessler2007}, and longitudinal and transverse spin relaxation of polarized neutrons and $^3$He~\cite{Serebrov2009,Pokotilovski2010,Petukhov2010,Fu2011}. A potential of the form $\hat{\sigma}\cdot\hat{r}$ can introduce a shift in the precession frequency of polarized particles in the presence of an unpolarized mass ~\cite{Youdin1996, Ni1999}, similar to that of a magnetic dipole in an external magnetic field, $\propto\vec{\mu}\cdot\vec{B}$. First measurement using this idea with $^3$He~\cite{Zheng2012} achieved a sensitivity of $5\times 10^{-3}$ Hz and restricted the coupling strength close to the current limit for force ranges from $10^{-4}$ to $10^{-2}$ m without any magnetic shielding. In this work, we present new results with a factor of 10-30 increase in sensitivity which constitute to our knowledge, the most stringent laboratory limit on $g_s g_p^n$ in the important ``axion window". The spin of the $^{3}$He is dominated by the spin of the neutron~\cite{Friar1990} so this result is directly interpretable in terms of the coupling $g_{p}^{n}$.  Constraints on T-odd and P-odd interactions of the $^{3}$He atom from bounds on the electric dipole moment (EDM) of its constituents from new physics at high-energy scales are highly suppressed due to the Schiff's screening~\cite{Schiff1963}, and the cancellation of the electronic EDM in the ground state of the $^3$He atom. Our work therefore also represents to our knowledge the most sensitive search for T-odd and P-odd interactions in the $^3$He atom at low energies.
\vspace{-0.02in}

\begin{figure}[h]
\centering
        \includegraphics[width=0.45\textwidth, height=0.31\textheight]{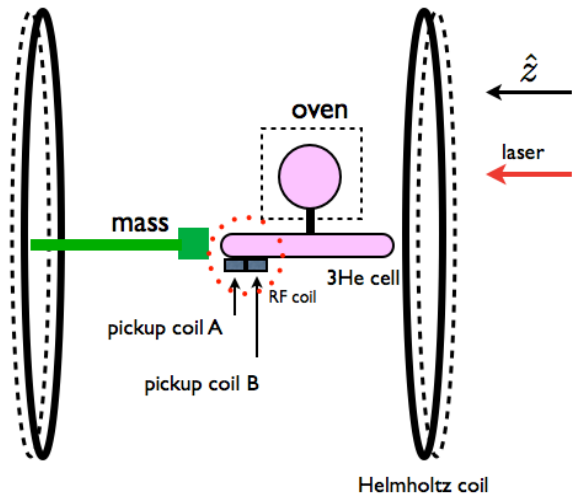}
	 \caption{Schematic of the SDSRF experiment (not to scale). The cylindrical polarized $^3$He cell is located in a uniform magnetic field. Correction coils (dashed-loop curves) compensate for residual holding field gradients. The direction of the laser and the holding field are along $\hat{z}$.}
        \label{fig:phaseII}
\end{figure}

The apparatus used in this work is based on the design of~\cite{Zheng2012}. We use a 7 amg high-pressure $^3$He cell as shown in Fig.~\ref{fig:phaseII}, which has an optical pumping chamber and a target chamber connected by a glass tube. $^3$He is polarized using spin-exchange optical pumping~\cite{Walker1997} in the spherical pumping chamber of  radius 4.3 cm. The polarized $^3$He atoms diffuse into the lower 40-cm long cylindrical chamber, which possesses two hemispherical glass windows at both ends with a thickness of 250 $\mu$m. 

We describe below a number of experimental improvements compared to~\cite{Zheng2012}, which enhance the sensitivity to SDSRF. A pair of  correction coils is applied to the Helmholtz coils to improve the uniformity of the holding field. Two identical pickup coils A and B of 2.5 cm diameter are located next to each other at the same end of the $^3$He cell. Pickup coil A is mounted below the window to measure the precession frequency shift of the polarized $^3$He nuclei due to SDSRF from the unpolarized mass. Pickup coil B is positioned to be insensitive to SDSRF; its signal is used to monitor the holding field drift. We subtract the frequencies measured in both coils and form $f_{A}^{\prime}=f_{A}-f_{B}$ for each measurement. The $^3$He cell position relative to the Helmholtz coils is adjusted to optimize the transverse spin relaxation time $T_2$ measured from coil A and B.

The holding field is tuned to produce a $^3$He Larmor frequency near 23.8 kHz. We apply a 24 kHz RF pulse to tip the spins by a small angle with negligible polarization loss. The polarized $^3$He nuclei induce EMFs in the pickup coils which are digitized and recorded. The precession frequency is determined first by applying a Fourier transform to a signal $s(t)$ in the time domain and obtaining the real and the imaginary parts of the signal in the frequency domain as $R(f)$ and $I(f)$, respectively. The Fourier transform is numerically calculated using Richardson extrapolation~\cite{Richardson1911}. The total amplitude is $S(f) = \sqrt{R^2(f)+I^2(f)}$. The reference frequency $f$ is then varied with a $10^{-6}$ Hz step to locate the maximum of $S(f)$, which is the precession frequency~\cite{Yan2012}.

Two samples are used as the unpolarized masses: a Macor ceramic mass block of dimensions $34\times 52\times38$ mm$^3$ used in~\cite{Zheng2012} and a liquid mixture of 1.02$\%$ MnCl$_2$ in pure water. These samples are chosen for their different nucleon densities, low magnetic impurities and magnetic susceptibilities, and minimal influence on the nuclear magnetic resonance (NMR) measurement procedure. The paramagnetic salt is added in order to compensate for the diamagnetism of the water. The magnetic susceptibility of this mixture is measured to be $<5\%$ of that of pure water. A stepping motor is used to move the ceramic mass a distance of 5 cm to 10 $\mu$m from the target chamber window. The salt water is stored in a cylindrical PTFE tank of a radius 34 mm and a length 37 mm.  Its one end is sealed with a 25 $\mu$m flexible PTFE film, which contacts the target chamber window. The other end of the liquid tank is connected to a flexible PTFE tube, which moves the liquid in and out of the tank using a nonmagnetic air cylinder actuated by a magnetically shielded switch.

In this work, we also improved the analysis method, which is described below. We define a mass-in state ($in$) or mass-out state ($out$) with the mass close to or away from the chamber window. Each measurement cycle employs two states of the mass position in the sequence ($in$, $out$, $in$, $out$) with a 60 second pause in the middle. We apply the analysis algorithm presented in~\cite{Swanson2010} to derive the frequency difference between the two states and remove any possible bias from linear or quadratic time-dependent frequency drifts. Assuming linear and quadratic time-dependent frequency drifts $f(t)\propto at + bt^2 \pm c$ with $a$ and $b$ being arbitrary constants, and $+c$($-c$) the frequency shift depending on the $in$($out$) state, the frequency difference between two successive cycles 1 and 2 is given by
\begin{align}
\Delta f &=\frac{1}{4}[ f_{in,1} - 3f_{out,1}+3f_{in,2}-f_{out,2}]\notag\\
&=\frac{1}{4}[(a\delta t+b\delta t^2+c )-3(a(2\delta t)+b(2\delta t)^2-c)\notag\\
&+3(a(3\delta t)+b(3\delta t)^2+c)-(a(4\delta t)+b(4\delta t)^2-c)]\notag\\
&= 2c
\end{align}
where $\delta t$ is the measurement time step (the time at the beginning of the first step of cycle 1 is taken as zero) and $f_{in/out, 1/2}$ is the frequency measured in the pickup coil A minus the pickup coil B for cycles 1 and 2, respectively. Higher-order algorithms produced the same results.

The mass in-mass out frequency difference can be measured in four different configurations of the apparatus corresponding to the directions of the main holding field and of the $^{3}$He polarization~\cite{Zheng2012}, each of which should possess the same magnitude of a frequency shift in the presence of a nonzero SDSRF proportional to the nucleon density of the mass. However, for our apparatus, two of these configurations possess residual field gradients in the sample large enough to lower the spin relaxation time $T_{2}$ and produce complicated line shapes whose frequency shifts, determined by a peak-finding algorithm of the type used in our analysis, are too sensitive to possible magnetic systematic effects induced by the mass. We therefore consider only two of the four configurations of the apparatus which lead to longer spin relaxation time $T_{2}$.

We define the precession frequency in different configurations of the holding field and the polarization direction as $f _{B,P}$ where the holding field $B=\pm$ and the polarization direction $P=\pm$.  The precession induced by SDSRF does not change after reversing the holding field, assuming the holding field rotation does not change the polarization of the $^3$He. But its precession direction now becomes opposite to the magnetic field-induced precession. Any systematic effects depending only on the polarization direction do not change after reversing the holding field. For different configuration, we can write
\begin{align}
f_{+ \pm} &= f_B\pm\Delta f_P+\Delta f_{\textrm{SDSRF}},\notag\\
f_{-\mp}  &= f_B\pm\Delta f_P-\Delta f_{\textrm{SDSRF}}
\end{align}
where $f_B$ is the magnetic field-dependent precession frequency including the holding field and the possible effect from the magnetic susceptibility of the mass, $\Delta f_P$ is the polarization-dependent frequency shift and $\Delta f_{\textrm{SDSRF}}$ represents the frequency shift due to SDSRF. In this work, the two configurations ($+-$) and ($-+$) which both have clean line shapes are considered in determining SDSRF
\begin{align}
\Delta f_{\textrm{SDSRF}}=\frac{1}{2}( \Delta f_{+-}- \Delta f_{-+})
\end{align}
where $\Delta f_{+-}$ and $\Delta f_{-+}$ are the frequency differences between the mass-in and mass-out states and the frequency measured in the pickup coil A minus the pickup coil B of each configuration.

\begin{figure}[t]
\centering
        \includegraphics[width=0.5\textwidth, height=0.3\textheight]{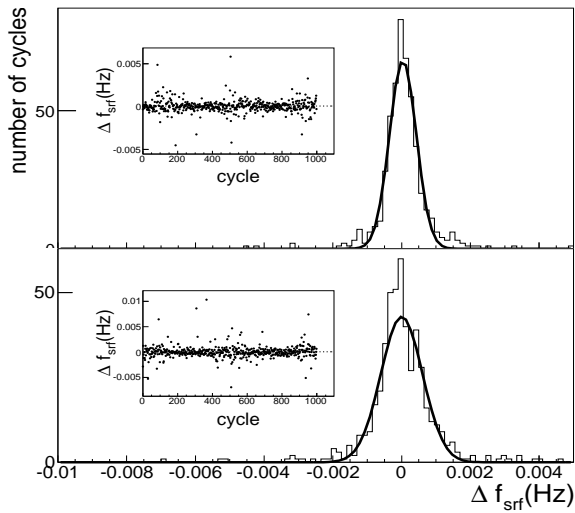}
	 \caption{Frequency difference of Macor ceramic(top) and salt water(bottom) for 1000 cycles. The average $\Delta f_{\textrm{srf}}$ is obtained by fitting the distribution as a Gaussian function. Small boxes show the $\Delta f_{\textrm{srf}}$ verse cycle. }
        \label{fig:data}
\end{figure}

We take 1000 cycles continuously for each configuration of each sample. The uncertainty in the measured frequency shift is given by
$\frac{1}{2}\sqrt{\sigma_{+-}^2+\sigma_{-+}^2}$. Fig.~\ref{fig:data}  and Table~\ref{tab:data} show the data of two samples and the average frequency shift due to SDSRF. The result shows that the average frequency difference of the salt water is consistent with zero.
\begin{table}[t]
		\centering
\begin{tabular}{|c|c|c|c|}
\hline
\backslashbox{Sample}{$\Delta f_{B, P}$}&$+~-$[$10^{-5}$Hz]&$-~+$[$10^{-5}$Hz]&$\Delta f_{\textrm{srf}}$[$10^{-5}$Hz]\\
\hline
ceramic&$0.6\pm1.3$&$-4.6\pm3.1$&$2.6\pm1.7$\\
\hline
salt water&$-3.3\pm 0.8$&$-1.7\pm 5.2$&$-0.8\pm2.6$ \\
\hline
\end{tabular}
		\caption{Data of each configuration of two samples. $\Delta f_{B,P}$ is the average frequency difference between mass-in and out states and the frequency measured in the pickup coil A minus the pickup coil B of each configuration.}
\label{tab:data}
\end{table}

Based on these results, we can constrain the force range and the coupling strength. The precession frequency shift due to SDSRF for each polarized $^3$He nucleus in the target chamber can be calculated by numerically integrating Eq.(\ref {eq:potential}) over the unpolarized mass as
\begin{align}
\Delta f(\vec{z}, \lambda, g_s g_p^n) =& \frac{2N}{2\pi\hbar}\int_{vol}V(\vec{r}-\vec{z})dr^3
\end{align}
where $N$ is the particle number density of the unpolarized mass, $vol$ is the total volume of the unpolarized mass and $\vec{z}$ is the distance from the surface of the mass to the polarized $^3$He nucleus. The precession signal measured by the pickup coil is $s(t) \propto \int (\vec{B}_{coil}\cdot \partial \vec{M}/\partial t) dr^3$ where $\vec{M}$ is the magnetization vector of $^3$He and $\vec{B}_{coil}$ is the field profile of the pickup coil which can be derived by using the reciprocity theorem~\cite{Insko1998}. The signal is written as
\begin{align}
s(t) &= C\{\int_{d} -B_x\cdot(f_0+\Delta f)\sin(2\pi(f_0 + \Delta f)t)\notag\\
&+ B_y\cdot (f_0+\Delta f)\cos{(2\pi(f_0 +\Delta f) t)dz}\}
\end{align}
where $f_0=\gamma B_0/2\pi$ is the Larmor frequency, $\gamma/2\pi = -3.24$ Hz/mG is the gyromagnetic ratio of $^3$He, $d$ is the thickness of the cell window and $C$ is a constant. Applying the Fourier transformation, the power spectrum of the signal can be calculated as,
\begin{align}
P(f^{\prime}) = C \{(&\int (f_0 + \Delta f) B_x \delta (f_0+ \Delta f- f^{\prime})dz)^2 \notag \\
+(&\int (f_0 + \Delta f) B_y \delta (f_0+ \Delta f - f^{\prime})dz)^2\}.
\end{align}
The average frequency observed by the pickup coil is 
\begin{align}
\bar{f}^{\prime} = \frac{\int f^\prime P(f^\prime)df^\prime}{\int P(f^\prime)df^\prime}.
\label{eq:deltaomega}
\end{align}
\begin{figure}[t]
\centering
        \includegraphics[width=0.48\textwidth, height=0.29\textheight]{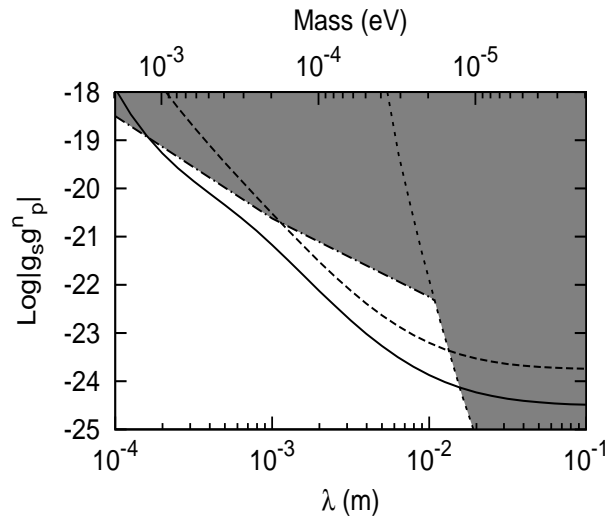}
	 \caption{Constraints on the coupling strength $g_s g_p^n$ as a function of the force range $\lambda$ and the equivalent mass of the ALPs. The dark gray area is the region excluded by previous works. The dotted curve is from~\cite{Youdin1996} and the dash-dotted curve is from~\cite{Petukhov2010}. The dashed (solid) curve is the constraint of the salt water (ceramic) sample within one standard deviation. }
        \label{fig:sensitivity}
\end{figure}

The frequency shift due to SDSRF is ${\Delta \bar{f}}(\lambda, g_s g_p^n) = \bar{f}^\prime - f_0$. Using the measured frequency difference in Table~\ref{tab:data}, the constraint on the coupling strength and the force range is determined as shown in Fig.~\ref{fig:sensitivity} where the dark gray area was ruled out by previous measurements. The dotted curve is from~\cite{Youdin1996} and the dash-dotted curve is from~\cite{Petukhov2010}. The dashed (solid) curve is the constraint of the salt water (ceramic sample) within one standard deviation. The measured frequency difference of the ceramic sample due to SDSRF is consistent with zero within 1.5 standard deviations. Our new results improve the constraint on SDSRF from the current limit in the range of $10^{-4}$ to $10^{-2}$ m by a factor of 10-30, which corresponds to a mass range of $2\times10^{-3}$ to $2\times10^{-5}$ eV for the pseudoscalar boson involved. This work represents the most sensitive search that sets a direct limit in the important ``axion window''.
\vspace{-0.02in}

Several methods can be employed in the future to further improve the sensitivity using polarized $^3$He. Obvious paths for improvement of the measurement include new magnetic holding field systems with better field uniformity and magnetic shielding, a smaller $^3$He cell with a lower pressure and thinner windows, unpolarized mass samples with higher fermion densities and lower magnetic susceptibilities, and a $^{129}$Xe comagnetometer. With these changes, we conclude that a factor of 10-100 improvement in the constraints of the coupling strength in the force range of $10^{-4}$ to $10^{-2}$ m is possible.
\vspace{-0.02in}

The authors thank M. Souza and T. Averett  for their help with the construction of the $^3$He cell. The authors also thank Y. Zhang, S. Jawalkar, T. Gentile, M.~Yu.~Khlopov and P.~Fayet for helpful discussions. This work was supported by the Duke University, the U. S. Department of Energy under Contract DE-FG02-03ER41231, and the U. S. National Science Foundation through grant PHY-1068712. K. Li, R. Khatiwada, E. Smith, M. Snow and H. Yan acknowledge support from the Indiana University Center for Spacetime Symmetries and A. Dennis from the IU STARS program.
\vspace{-0.25in}


\end{document}